# InfoFlowNet: A Multi-head Attention-based Self-supervised Learning Model with Surrogate Approach for Uncovering Brain Effective Connectivity

Chun-Hsiang Chuang[a,b,c*], Shao-Xun Fang[a,d], Chih-Sheng Huang[e,f,g], Weiping Ding[h]

[a] Research Center for Education and Mind Sciences, College of Education, National Tsing Hua University, Hsinchu, Taiwan

[b] Institute of Information Systems and Applications, College of Electrical Engineering and Computer Science, National Tsing Hua University, Hsinchu, Taiwan

[c] Department of Education and Learning Technology, National Tsing Hua University, Hsinchu, Taiwan

[d] Department of Computer Science and Engineering, National Taiwan Ocean University, Keelung, Taiwan.

[e] Department of Artificial Intelligence Research and Development, Elan Microelectronics Corporation, Hsinchu, Taiwan

[f] College of Artificial Intelligence and Green Energy, National Yang Ming Chiao Tung University, Hsinchu, Taiwan

[g] College of Electrical Engineering and Computer Science, National Taipei University of Technology, Hsinchu, Taiwan

[h] School of Information Science and Technology, Nantong University, Nantong, China

* Correspondence should be addressed to the following:

Chun-Hsiang Chuang,
Deputy Director, Center for Education and Mind Sciences, College of Education, National Tsing Hua University (NTHU), Taiwan
Associate Professor, Institute of Information Systems and Applications, College of Electrical Engineering and Computer Science, NTHU
Treasure, Taipei Chapter of IEEE Computational Intelligence Society

Phone: +886 (03) 5715131#78607
E-mail: ch.chuang@ieee.org

# ABSTRACT

Deciphering brain network topology can enhance the depth of neuroscientific knowledge and facilitate the development of neural engineering methods. Effective connectivity, a measure of brain network dynamics, is particularly useful for investigating the directional influences among different brain regions. In this study, we introduce a novel brain causal inference model named InfoFlowNet, which leverages the self-attention mechanism to capture associations among electroencephalogram (EEG) time series. The proposed method estimates the magnitude of directional information flow (dIF) among EEG processes by measuring the loss of model inference resulting from the shuffling of the time order of the original time series. To evaluate the feasibility of InfoFlowNet, we conducted experiments using a synthetic time series and two EEG datasets. The results demonstrate that InfoFlowNet can extract time-varying causal relationships among processes, reflected in the fluctuation of dIF values. Compared with the Granger causality model and temporal causal discovery framework, InfoFlowNet can identify more significant causal edges underlying EEG processes while maintaining an acceptable computation time. Our work demonstrates the potential of InfoFlowNet for analyzing effective connectivity in EEG data. The findings highlight the importance of effective connectivity in understanding the complex dynamics of the brain network.

# 1. INTRODUCTION

The study of brain signatures holds great potential for advancing our understanding of neuroscience and the development of neural engineering techniques. Rather than solely focusing on the activity of individual brain regions, researchers have shifted their attention toward brain networks [1, 2] to gain a more comprehensive depiction of the couplings and associations among distant brain regions. Accurately representing brain network topology is crucial for thoroughly investigating brain function and behavior complexities. This necessitates the creation of pragmatic models capable of capturing the intricate interactions among various brain regions.

Brain networks can be categorized into three primary types: anatomical, functional, and effective connectivity. Anatomical connectivity provides the structural basis of physically connected brain elements [3]. In contrast, functional connectivity quantifies the dependence among interacting brain regions responsible for specific tasks or behaviors [4]. Various connectivity measures based on statistical models or signal similarity have been proposed. Commonly used functional connectivity measures include Pearson's correlation coefficient [5], coherence [6], and phase locking value, which assess synchronization and symmetry among brain regions, resulting in a non-directional connectivity representation of a brain network. Effective connectivity, also known as directed functional connectivity, models the causal relationship among brain regions [1, 7]. Strategies such as Granger causality analysis [8], dynamic causal modeling [9], and transfer entropy [10] have been designed to investigate asymmetric or directional influences among brain regions. The strength and direction of information flow are visualized in an arrow diagram or circular graph with bidirectional or unidirectional edges between nodes. In recent decades, significant strides have been made in applying these connectivity measures to diverse neurophysiological data, leading to the identification of complex brain networks spanning a wide range of research topics, from human cognitive states to disease diagnosis [11-13].

Recent studies on brain networks have transitioned from examining neural associations to exploring causation [14]. Within a causal inference framework, brain connectivity analysis involves selecting specific brain regions and examining the directionality, existence, and strength of connections between them. For instance, the Granger causality model (GCM)[8] applies a multivariate autoregressive model [15] to electroencephalogram (EEG) data to determine causal relationships. Signal A is considered to cause Signal B if the past information of Signal A can predict the current information of Signal B. This causal interaction between signals is defined using a multivariate autoregressive model, which can be estimated through algorithms such as ARfit [16] or Kalman filtering [17]. Three tests are typically used to validate the model fitness: whiteness test of the model residuals, consistency tests, and stability tests. The magnitude of causality among pairs of signals is determined by comparing the residual values of unrestricted and restricted models [18]. Here, the unrestricted model is a multivariate autoregressive model that includes all available signals. In contrast, the restricted model omits certain signals. If the exclusion of Signal A results in increased residual values when predicting Signal B, indicating that the current information of Signal B cannot be accurately predicted without the past information of Signal A, then Signal A is deemed to cause Signal B, establishing a source–sink relationship.

Numerous deep-learning-based approaches have been developed to generate interpretable graphical representations of brain signals. For instance, a diffusion convolutional recurrent neural network [19] uses encoders, decoders, and gated recurrent units to capture the temporal evolution of brain signals. BrainGNN (where GNN represents a graph neural network)[20] seeks to extract a graphical representation that is most informative for classification tasks. The use of brain graph features has been shown to outperform conventional models in training classifiers [21, 22]. Additionally, deep convolutional neural network models, such as the temporal causal discovery framework (TCDF)[23], have been developed to uncover causal relationships between time series. The TCDF uses an attention mechanism and the estimated attention scores to interpret causal relationships among time series. Furthermore, the receptive field (RF), influenced by the kernel size and number of hidden layers, helps accommodate the time delay between sources and sinks. However, the performance of the TCDF deteriorates with limited time series lengths, rendering it less effective for capturing phasic changes in brain network dynamics.

Therefore, this paper proposes InfoFlowNet, a novel model aimed at efficiently capturing causal interactions among brain processes. InfoFlowNet leverages multi-head attention [24] and self-supervised learning to learn associations between time series, enabling nonlinear prediction. This deep learning approach reduces dependence on the linear autoregressive model, model order selection, and model fitting algorithm used in the GCM. Inspired by the unrestricted and restricted model comparison in the GCM, this study introduces a simple causal magnitude estimation strategy by incorporating surrogate data into the inference process. Rather than drawing causal inferences through attention scores, as in the TCDF, the proposed method assesses the causal strength of a specific process by measuring changes in model prediction errors when its time series is randomly shuffled. The experimental results provide compelling evidence of the superior performance of InfoFlowNet compared with existing frameworks.

**2. InfoFlowNet**

The InfoFlowNet model is designed to uncover effective connectivity in brain processes. Leveraging a multi-head attention mechanism (Section 2.1), InfoFlowNet is trained to capture causal relationships among brain processes. Within the context of a self-supervised learning framework, using a similarity-based loss function, the proposed model operates on the assumption that if a causal relationship exists among various processes, one process can be inferred or predicted by the remaining processes. Furthermore, a new causal inference method is introduced to ascertain the causal strength of one process on another (Section 2.2). The robustness and feasibility of the proposed InfoFlowNet in capturing phasic changes in effective connectivity are assessed by conducting experiments using one simulated dataset and two EEG datasets (Section 3).

*2.1 Network Architecture*

The network architecture of InfoFlowNet is shown in Fig. 1A. Given an input process selected from predefined windows of interest, $\mathbf{X}^{\dagger} \in \mathbb{R}^{d \times T}$, consisting of $d$ channels and $T$ sample points, the first feature map, $\mathbf{H} \in \mathbb{R}^{d \times T}$, in the hidden layer is extracted using two 1D-convolution (denoted as Conv1d) layers. Each convolution operation is followed by batch normalization (denoted as BatchNorm1D). The RF [25] of the convolution operation determines the number of sample points in the input process that the feature map extracts. As illustrated in Fig. 1B, considering a stride of 1 and padding of $\frac{(\text{ks}-1)}{2}$, the RF size, $r$, can be obtained as follows:

$$r = \ell \times \text{ks} - 1, \tag{1}$$

where $\ell$ represents the number of convolution layers, and $\text{ks}$ denotes the kernel size. Note that $r$ is adjusted to accommodate for the time delay inherent in the potential causality among processes.

Subsequently, InfoFlowNet integrates temporal encoder/decoder and multi-head attention layers to augment its capacity to discern both temporal and spatial relationships among brain processes. The resulting second feature map, denoted as $\mathbf{H}' \in \mathbb{R}^{d \times T}$, facilitates the reconstruction of causal interactions within and between brain processes over time.

Specifically, the feature map $\mathbf{H}$ is encoded into a $d \times T_e$ representation of processes through a fully connected layer, capturing the temporal dependencies shared among the processes, for use in subsequent attention layers. Next, as shown in Fig. 1C, we first segment $\mathbf{H}_e$ into $h$ equal-dimensional sets $\in \mathbb{R}^{d \times (T_e/h)}$, which are then linearly transformed into sets of query, key, and value matrices, $\{\mathbf{Q}_t, \mathbf{K}_t, \mathbf{V}_t \in \mathbb{R}^{d \times (T_e/h)} | t = 1, 2, \ldots, h\}$, using weight matrices $\{\mathbf{W}_t^Q, \mathbf{W}_t^K, \mathbf{W}_t^V \in \mathbb{R}^{d \times (T_e/h)} | t = 1, 2, \ldots, h\}$. The dot product of $\mathbf{Q}_t$ and $\mathbf{K}_t^\intercal$, representing the attention score matrix $\mathbf{A}_t \in \mathbb{R}^{d \times d}$, depicts the similarity between the $t^{th}$ key and query. To ensure that $h$ attention score matrices are scaled appropriately and are interpretable across heads, a normalized attention score matrix $\mathbf{A}_t' \in \mathbb{R}^{d \times d}$ is obtained using a softmax function. To better uncover the causal relationships between processes while mitigating the impact of autoregression within each process, a masking approach is used to disregard the diagonal values in $\mathbf{A}_t$. This strategy enhances the capability of InfoFlowNet in reinforcing the causality among processes. The final step is to concatenate $h$ dot-products of $\mathbf{A}'_t$ and $\mathbf{V}_t$ to yield $\mathbf{H}'_t \in \mathbb{R}^{d \times T_e}$. Then, a temporal decoder layer is applied to obtain feature maps $\mathbf{H}' \in \mathbb{R}^{d \times T}$, which are mapped to the final output $\mathbf{X}^\ddagger \in \mathbb{R}^{d \times T}$ through a Conv1d layer.

InfoFlowNet is trained using self-supervised learning approach to minimize the discrepancy between the actual and predicted processes. Considering both the amplitude and phase of the processes, the loss function is a combination of the mean squared error (MSE) and cosine similarity between the actual processes $\mathbf{X}^\dagger$ and predicted processes $\mathbf{X}^\ddagger$, expressed as $\text{Loss} = \text{mse}(\mathbf{X}^\dagger, \mathbf{X}^\ddagger) + \left(1 - \cos(\mathbf{X}^\dagger, \mathbf{X}^\ddagger)\right)$. In this study, the kernel size is configured to $\text{ks} = 15$, $T_e$ is set as 512, the number of training epochs is 100, and the batch size is 128.

*2.2 Model Inference*

The central concept of InfoFlowNet is to estimate causal influence by interpreting model loss. This model is designed to recover each process with minimal loss when all input processes are available. Conversely, the absence of a key process crucial to others may result in inaccurate predictions. The resulting loss is then used to quantify the causal influence. To this end, we introduce a novel method (Fig. 2) to assess the causal strength between processes.

Consider a time series with $d$ channels, segmented into $k$ windows of equal length $T$. Effective connectivity is assumed to be present around specific windows. These windows are defined as the windows of interest (WOIs), and the corresponding data segments are used to train InfoFlowNet. After the training phase (as detailed in Section 2.1), each window's data segment undergoes the model inference phase. Subsequently, the similarity between the input and output processes is calculated.

For instance, consider the $w^{\text{th}}$ window, where $w = 1, 2, \ldots, k$. Its output process, $\mathbf{X}'(w)$, predicted by InfoFlowNet, is compared with the input process, $\mathbf{X}(w)$, to obtain the similarity between them, denoted as $\Delta(\mathbf{X}(w), \mathbf{X}'(w)) \in \mathbb{R}^d$. It is hypothesized that the similarity is high if the intrinsic connectivity network among processes of the $w^{\text{th}}$ window is similar to that of the WOI.

Simultaneously, $d$ randomly shuffled surrogates of the $w^{\text{th}}$ window are generated by simply permuting the time course of the $i^{\text{th}}$ process in random order, where $i = 1, 2, \ldots, d$. The model inference step is then repeated on the surrogate, $\widetilde{\mathbf{X}}_i(w)$, and produces its predicted output, $\widetilde{\mathbf{X}}'_i(w)$, followed by the similarity estimation for these two processes to yield $\Delta\left(\widetilde{\mathbf{X}}_i(w), \widetilde{\mathbf{X}}'_i(w)\right) \in \mathbb{R}^d$. Due to the shuffling, the distortion of the $i^{\text{th}}$ process may lead to certain differences between $\Delta(\mathbf{X}(w), \mathbf{X}'(w))$ and $\Delta\left(\widetilde{\mathbf{X}}_i(w), \widetilde{\mathbf{X}}'_i(w)\right)$. It is hypothesized that this difference functions as an indicator of the influence of the $i^{\text{th}}$ process on the remaining processes.

If the $i^{\text{th}}$ process causes another process, shuffling the $i^{\text{th}}$ process might obstruct the model from accurately predicting that process. Conversely, if the $i^{\text{th}}$ process does not cause that process, the model might still predict the process even if the $i^{\text{th}}$ process is shuffled. Therefore, the cost of the random shuffle, i.e., the array elements of the difference between $\Delta(\mathbf{X}(w), \mathbf{X}'(w))$ and $\Delta\left(\mathbf{X}(w), \widetilde{\mathbf{X}}'_i(w)\right)$, is considered to define the information flow magnitude from the $i^{\text{th}}$ source process to all $d$ sink processes at the $w^{\text{th}}$ window. This value is denoted as $\mathbf{dIF}_{i\bullet}(w) \in \mathbb{R}^d$, where $i =$ $1, 2, \ldots, d$. Commonly used similarity measures such as correlation, cross-correlation, and cosine similarity can be used to gauge the difference between actual and predicted processes.

*2.3 Information Flow Magnitude*

For the $w^{\text{th}}$ window, the $\mathbf{dIF}_{ij}(w) \in \mathbf{dIF}_{i\bullet}(w)$, $j = 1, 2, \ldots, d$, can be calculated using the following steps:

(1) Apply Model Inference: Model inference is performed on both the original windowed data and their surrogates to obtain the corresponding predicted outputs. Subsequently, a straightforward measure is used to gauge the similarity between them, applying commonly used similarity measures, such as the MSE, Pearson's correlation coefficient, cross-correlation, or dynamic time warping. The resulting arrays, $\Delta(\mathbf{X}(w), \mathbf{X}'(w)) \in \mathbb{R}^d$ and $\Delta\left(\mathbf{X}(w), \widetilde{\mathbf{X}}'_i(w)\right) \in \mathbb{R}^d$, contain $d$ similarity values for each channel. Higher values denote greater similarity between the input/output pairs. If using a distance-based measure such as the MSE, its value is inverted to ensure consistency with other similarity measures. For correlation-based measures, any negative correlation coefficient is regarded as an invalid prediction result, and its similarity score is set as 0. The maximum value of $\Delta\left(\mathbf{X}(w), \widetilde{\mathbf{X}}'_i(w)\right)$ to $\Delta(\mathbf{X}(w), \mathbf{X}'(w))$ is limited, because $\mathbf{X}'(w)$ is expected to approximate $\mathbf{X}(w)$ more closely than $\widetilde{\mathbf{X}}'_i(w)$. This can be expressed as

$$f\left(\Delta\left(\mathbf{X}(w), \widetilde{\mathbf{X}}'_i(w)\right)\right)$$

$$= \begin{cases} \Delta\big(\mathbf{X}(w), \widetilde{\mathbf{X}}'_i(w)\big), & \text{if } \Delta\big(\mathbf{X}(w), \widetilde{\mathbf{X}}'_i(w)\big) < \Delta(\mathbf{X}(j), \mathbf{X}'(j)) \\ \Delta(\mathbf{X}(w), \mathbf{X}'(w)), & \text{otherwise} \end{cases} \quad (2)$$

In this study, we compare the performance of various similarity measures, specifically correlation (denoted as corr), cross-correlation (denoted as x-corr), and cosine similarity.

(2) Quantify Shuffle Cost: The difference between similarities is calculated to quantify the cost of random shuffling:

$$\boldsymbol{c}_i(w) = f(\Delta(\mathbf{X}(w), \mathbf{X}'(w))) - f\left(\Delta\big(\mathbf{X}(w), \widetilde{\mathbf{X}}'_i(w)\big)\right), \quad (3)$$

where $\boldsymbol{c}_i(w) = \{c_{i1}(w), c_{i2}(w), \ldots, c_{id}(w)\} \in \mathbb{R}^d$ denotes the cost of similarity resulting from shuffling the $i^{\text{th}}$ process in the $w^{\text{th}}$ time window. Note that the cost is a non-negative value, ranging from 0 to $\Delta(\mathbf{X}(w), \mathbf{X}'(w))$.

(3) Normalization: Normalization is applied to each $c_{ij}(w) \in \boldsymbol{c}_i(w)$ to obtain the final information flow magnitude. This ensures that the causality from the $i^{\text{th}}$ source to the $j^{\text{th}}$ process, $\mathbf{dIF}_{ij}(w)$, ranges between 0 and 1. This transformation is achieved using a modified sigmoid function mapping:

$$\mathbf{dIF}_{ij}(w) = g(c_{ij}(w)) = \left[\left(\frac{1}{1 + e^{-\lambda \times c_{ij}(w)}}\right) - 0.5\right]/0.5, \quad (4)$$

which can be simplified as

$$\mathbf{dIF}_{ij}(w) = g(c_{ij}(w)) = \frac{1}{4}\left(\frac{1 - e^{-\lambda \times c_{ij}(w)}}{1 + e^{-\lambda \times c_{ij}(w)}}\right). \quad (5)$$

where $\lambda$ is the scaling parameter.

The complete algorithm of the proposed model is presented in Fig. 3.

## 3. EXPERIMENTS AND RESULTS

*3.1. Datasets*

The performance of InfoFlowNet, was validated through one simulated dataset and two EEG datasets, outlined in the following text.

The simulated dataset included 100 samples, each consisting of three distinct synthetic time-series processes: sine, sawtooth, and random waves. Each wave constituted a 500-point process. As illustrated in Fig. 4A, the sawtooth process was configured to causally influence sine and random processes from time points 151 to 350, with the causal magnitude set as 0.5. Each sample was segmented into multiple 100-point windows, with each consecutive window overlapping by 99 points, resulting in 100 × 401 windows. Among these windows, those falling within time points 151 to 350, designated as the WOI, were used to train the InfoFlowNet.

Two EEG datasets were drawn, one from a psychological vigilance task and the other from a sustained-attention driving experiment[26, 27]. The experiments were designed to evaluate participants' vigilance by gauging their reaction time to randomly presented visual stimuli and vehicle departure events. The psychological vigilance and sustained-attention driving experiments involved 1,080 and 581 samples, respectively, with sampling performed at a rate of 500 Hz. Guided by a hypothesis aimed at exploring the effective connectivity associated with these responses, EEG processes from the fronto-parietal network of the brain [28], specifically the midline area (including Fz, Cz, Pz, and Oz), were analyzed. Each sample was further segmented into multiple 100-point windows, each overlapping by 10 points, resulting in a total of 141 windows per sample. The WOIs were determined around the average reaction times, which were 370.8 ± 181.1 ms and 948.9 ± 306.7 ms for the two tasks. Specifically, four and eight windows were selected for the tasks to encompass the average reaction time, extending to one standard deviation above and below, respectively.

The connectivity graphs and causality values obtained through InfoFlowNet were compared with those derived from the GCM and the TCDF[23]. This comparison was visualized through a network diagram, i.e., a circular graph, with directional edges representing the statistically significant information flows from source to sink processes.

*3.2. Simulation Results*

Figure 5 shows the temporal variations in the **dIF** values between simulated source and sink processes, determined by InfoFlowNet, in three distinct scenarios: without self-attention, with single-head attention, and with multi-head attention (number of heads = 8) mechanisms. The three color traces in the figure represent the three matrices used in the similarity measurement. The nearly flat **dIF** changes observed in the off-diagonal subfigures reveal that the model lacking a self-attention mechanism (Fig. 5A) failed to capture the causality between both the sawtooth and sine waves as well as that between the sawtooth and random waves. This model tended to focus exclusively on the temporal relationships inherent in the processes' own signal sequences. In contrast, the self-attention mechanism enabled the model to capture the causality among processes, as demonstrated in Figs. 5B and 5C, evidenced by the increased **dIF** values in the source–sink pairs of sawtooth→sine and sawtooth→random, between time points 151 and 350. However, the model using the single-head attention mechanism (Fig. 5B) detected some spurious causality in the source–sink pairs of sine→sawtooth, sine→random, random→sine, and random→sawtooth. Overall, the model using the multi-head attention mechanism (Fig. 5C) outperformed the other two models, successfully capturing the simulated causality, as anticipated.

Additionally, the choice of the similarity metric significantly influenced the model performance, with correlation and cosine (blue and green traces, respectively) outperforming cross-correlation (red trace). For instance, spurious causality was observed in the source–sink pairs of random→sine, as reflected by the increased **dIF** values. This phenomenon was observed only in the case of models using both single- and multi-head attention mechanisms with cross-correlation, and not with correlation and cosine similarity measures.

Figure 6 compares the time-varying causality changes between source and sink processes, as estimated by the GCM, the TCDF, and InfoFlowNet. InfoFlowNet outperformed the other two methods. While the GCM and the TCDF failed to capture the causality of sawtooth→random and sawtooth→sine, respectively, both methods erroneously detected a spurious causality of sine→sawtooth.

To mitigate the influence of self-causality of processes on the **dIF**, a masking mechanism was introduced. This mechanism disregarded the effects of diagonal values in the attention score matrix, thereby allowing the model to effectively learn the relationships between processes. Figure 7 compares the differences in **dIF** estimation with and without the use of the masking mechanism in InfoFlowNet. The values estimated using the masking mechanism aligned more closely with expectations, demonstrating that InfoFlowNet was capable of capturing the causal relationships between sawtooth→sine and sawtooth→random. However, this model tended to incorrectly identify causal relationships between sine→sawtooth and sine→random, likely attributable to the excessive similarity in the sine and sawtooth waveforms.

*3.3. Real-world EEG Experiments and Results*

*3.3.1   Psychomotor vigilance task*

Figure 8A demonstrates the effectiveness of InfoFlowNet in capturing the effective connectivity between real EEG signals. With the number of heads set to eight and the similarity measured using the cosine measure, InfoFlowNet could capture causality variations over time and events. The influence of masking was notably visible in the non-diagonal connections, where the **dIF** significantly increased with the masking mechanism.

Figure 8B compares the causal relationships modeled by three methods: the GCM, the TCDF, and InfoFlowNet. The chordal graph displays the event-related causal relationships between EEG processes, with each edge representing a significant difference in **dIF** between the WOI and corresponding baseline. InfoFlowNet could identify a greater number of causal relationship features compared with the GCM and the TCDF, which only identified causal relationships from Cz→Fz and Cz/Pz→Fz, respectively.

*3.3.2   Driving task*

InfoFlowNet was applied to a publicly accessible dataset of lane-keeping driving data to investigate the brain's effective connectivity in response to an unexpected lane departure event. As shown in Fig. 9A, the WOI (highlighted in purple) was selected for training the model and examining the causal relationships between brain regions in response to the event.

In the original experiment involving reaction time (RT) and sustained-attention tasks, each trial was paired with an RT value. The experiment was designed to test the hypothesis that effective connectivity may vary with changes in the task performance of a participant. To this end, the baseline effective connectivity of different task performances was estimated and compared. Specifically, as shown in Fig. 9B, trials with RTs of 5%–15% and 85%–95% were assigned to the optimal performance group ($G_{opt}$) and suboptimal performance group ($G_{subopt}$), respectively. Each group involved 54 trials. The mean RTs for $G_{opt}$ and $G_{subopt}$ were 715.7 ± 11.7 ms and 1425.5 ± 87.8 ms, respectively. Figure 9C illustrates the differences in **dIF** between the two groups. The **dIF** from Fz to Pz/Oz, Pz to all channels, and Oz to Cz/Pz in the $G_{subopt}$ group showed significant differences compared with those in the $G_{opt}$ group. This result suggested that as the participants' RTs to the events increased, the causal strength of Fz, Pz, and Oz decreased. In addition, most causal relationships involving Cz did not show significant changes in relation to task performance.

*3.4 Computation time*

Table 1 presents the computation time (in seconds) required by the GCM, the TCDF, and InfoFlowNet to estimate effective connectivity features on a dataset with dimensions of 4 channels × 100 points × 100 trials. Among the three methods, the GCM demonstrated the highest efficiency, with the shortest computation time. As anticipated, InfoFlowNet required the longest duration, while the TCDF fell in the intermediate range.

## 4. DISCUSSION

This study introduces InfoFlowNet as an effective method for capturing causal interactions among brain processes. This novel deep learning approach enables nonlinear prediction, reducing the reliance on the linear autoregressive model. The proposed surrogate approach can estimate the causal strength of specific processes, revealing complex associations between source and sink processes. In the ensuing discussion, we delve deeper into three critical aspects of the functionality and efficacy of InfoFlowNet: 1) Verification of the effectiveness of InfoFlowNet in signal reconstruction through the analysis of residuals; 2) exploration of how InfoFlowNet's multi-head attention and masking mechanisms impact the reconstruction of processes; and 3) formulation of statistical inferences regarding the proposed **dIF** among brain processes. These focal points are integral to understanding the capabilities and limitations of InfoFlowNet, offering insights into its potential applications and areas for future development in the study of brain connectivity and signal processing.

*4.1 Distribution of Model Errors*

Reliable interpretation of a model's predictions and inferences is crucial for gaining a deeper understanding of the underlying processes being studied. One approach to assess the validity of a model involves closely examining the errors between the original and predicted processes, commonly referred to as residuals. Random residuals are always desirable, indicating that the model's inferences are not influenced by unaccounted-for factors. However, the presence of patterns or structure in the errors suggests systematic deviations from randomness, indicating the need for model refinement to capture the intricacies and temporal dependencies in the data.

Whiteness tests, such as the Ljung–Box test or the Durbin–Watson test, can be used to assess whether the errors between the original and predicted processes exhibit serial correlation or significant departures from randomness. As demonstrated in Fig. 10, these whiteness tests were applied to the prediction errors of the InfoFlowNet model, using the dataset from the psychomotor vigilance task experiment. The figure is divided into four subfigures, representing the error distribution, time-series plot of errors, autocorrelation function plot, and p-values obtained from autocorrelation tests. Collectively, the errors portrayed a normal distribution with low and statistically insignificant autocorrelation, indicating randomness in the errors. In general, if the whiteness test is violated, additional model refinement may be necessary to enhance its accuracy and robustness. Model training should not focus exclusively on minimizing errors, but also on pursuing whiteness in the residuals.

*4.2 Multi-head Attention and Masking Mechanisms*

InfoFlowNet functions as a self-supervised learning model specialized in signal reconstruction. Once trained, the model uses the similarity between the predicted and original processes, along with a surrogate approach, to determine the presence and intensity of causal relationships. Thus, comprehending the model's methodology for reconstructing signals is essential for its effective application in subsequent analyses. Figure 11 highlights the crucial role of multi-head attention in InfoFlowNet for adept signal reconstruction. In the absence of this attention mechanism, InfoFlowNet encountered challenges in accurately predicting the patterns of Fz. In contrast, the integration of an attention layer significantly enhanced the model's reconstruction capabilities, a benefit that was further amplified with the increase in the number of heads.

The inclusion of the masking mechanism within InfoFlowNet significantly influenced the causal relationships involved in the mutual reconstruction of signals. Figure 12 illustrates this phenomenon under a configuration of eight heads without masking. The attention scores were predominantly self-directed, focusing mainly on their respective channels. This approach proved ineffective in facilitating the model's learning of causal relationships between processes. In contrast, the introduction of masking enabled the model to divert its focus from its own signal, resulting in more evenly distributed attention scores that were not concentrated on the diagonal. This shift not only enhanced the model's capacity to discern inter-process causalities but also facilitated the visualization of process dependencies, as evidenced by the varied values across the attention score matrix for each head.

*4.3 Statistical Inferences*

Three critical statistical inferences regarding the causality among brain processes [15] could be formulated: 1) Is the **dIF** a non-zero value? 2) Does the **dIF** fluctuate over time or events? 3) Is there a difference in **dIF** between various conditions?

The first question can be addressed through a surrogate statistical test for non-zero values. Similar to the phase randomization technique [29], the initial step involves creating a null distribution of the estimator for a scenario with no information flow. The shuffling procedure (Section 2.2) can be repeatedly executed to establish the null distribution. Subsequently, a confidence interval is defined to test whether the **dIF** value significantly deviates from the null distribution at a specific confidence level.

The second question can be addressed by applying the proposed method using a sliding-window approach across experimental trials that are time-locked to specific stimuli or events. Grounded in domain knowledge, it is feasible to hypothesize the existence of possible causality among brain regions within certain time windows. As depicted in Fig. 2, after calculating the **dIF** matrices for all time windows, an event-related analysis may be conducted to uncover the time-varying **dIF** dynamics between each pair of brain processes. Given a sufficient number of trials, paired-sample or non-parametric statistical analyses can be performed to compare the **dIF** of the selected time window against that of the baseline.

The third question pertains to exploring whether the brain network changes in response to different experimental treatments. This investigation involves two groups: the experimental group and the control group. Data from the experimental group are used to train InfoFlowNet. Subsequently, the inference process is applied to both groups to calculate the **dIF**. An independent sample statistical analysis is then conducted to examine the differences in **dIF** between the groups.

## 5. CONCLUSIONS

This study was aimed at developing InfoFlowNet, a novel self-supervised learning model for brain causal inference, which leverages a multi-head attention mechanism and a surrogate approach to estimate the directional flow of information among EEG time series. The effectiveness of InfoFlowNet in extracting time-varying causal relationships was demonstrated, as evidenced by fluctuations in **dIF** values. A comparative analysis highlighted that InfoFlowNet outperforms other causal inference models in identifying significant causal edges within EEG processes, while still maintaining an acceptable level of computational efficiency in causal inference. These findings highlight the potential of InfoFlowNet as a tool for analyzing effective connectivity in EEG research.

**Acknowledgments**

This work was supported by the Ministry of Science and Technology, Taiwan (project numbers: MOST 111-2636-E-007-020, 110-2636-E-007-018, 110-2221-E-019-052-MY3); National Science and Technology Council, Taiwan (project number: NSTC 112-2636-E-007-009); Research Center for Education and Mind Sciences, National Tsing Hua University; and Yin Shu-Tien Educational Foundation.

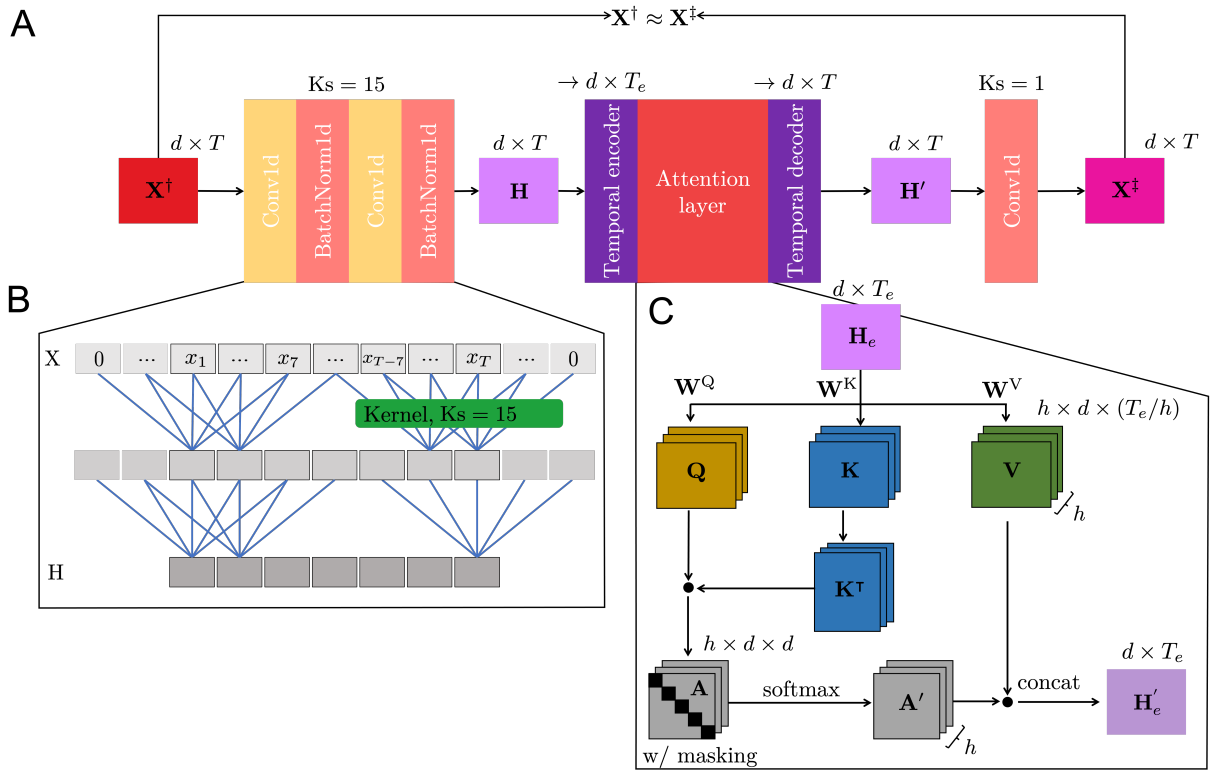

**Figure 1.** InfloFlowNet model. (A) Model architecture, including three 1D-convolution, temporal encoder/decoder and multi-head attention layers. (B) Receptive field. (C) Multi-head attention layer.

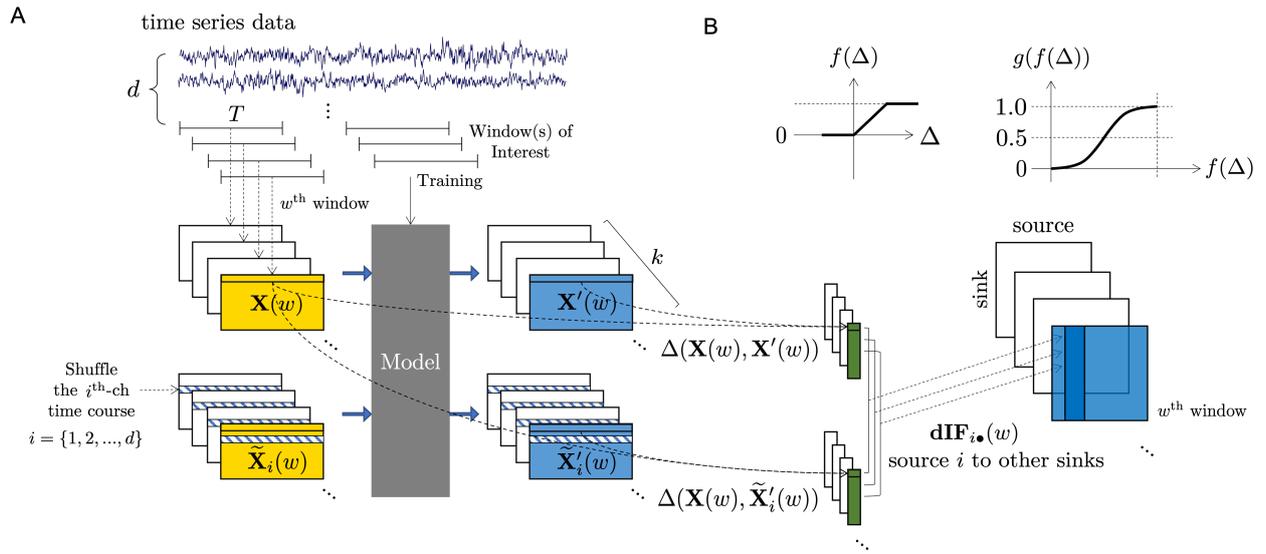

**Figure 2.** Estimation of directional information flow (dIF) magnitude. (A) Model training (B) Causal inference through shuffled surrogates.

```
Algorithm: dIF calculation
```
---

**Input: X**

**Output:** $\mathrm{dIF}_{i\bullet}(w) \in \mathbb{R}^d$, the causality from the $i^{\mathrm{th}}$ source to all sinks at the $w^{\mathrm{th}}$-window

**Initialization:**

Divide **X** into $k$ windows

Define $s \leq k$ windows as WOIs, denoted as $\mathbf{X}^\dagger \in \mathbb{R}^{d \times \mathrm{T}} \subset \mathbf{X}$

**InfoFlowNet modeling:**

    **for** each epoch **do**

        **for** each mini-batch **do**

            $\mathbf{H} \leftarrow \mathrm{2layerConv1d}(\mathbf{X}^\dagger)$

            $\mathbf{H}_{\mathrm{e}} \leftarrow \mathrm{temporal\ encoder}(\mathbf{H})$

            **for** $t = 1;\ t \leq h$ **do**

                $\mathbf{Q}_t \leftarrow \mathbf{W}_t^{\mathrm{Q}} \cdot \mathbf{H}_{\mathrm{e}}, \mathbf{K}_t \leftarrow \mathbf{W}_t^{\mathrm{K}} \cdot \mathbf{H}_{\mathrm{e}}, \mathbf{V}_t \leftarrow \mathbf{W}_t^{\mathrm{V}} \cdot \mathbf{H}_{\mathrm{e}}$

                $\mathbf{A}'_t \leftarrow \mathrm{softmax}(\mathrm{masking}(\mathbf{A}_t \leftarrow \mathbf{Q}_t \cdot \mathbf{K}_t^\top))$

            **end for**

            $\mathbf{H}'_{\mathrm{e}} = \mathrm{concat}(\mathbf{A}'_1 \cdot \mathbf{V}_1, \mathbf{A}'_2 \cdot \mathbf{V}_2, \ldots, \mathbf{A}'_h \cdot \mathbf{V}_h)$

            $\mathbf{H}' \leftarrow \mathrm{temporal\ decoder}(\mathbf{H}'_{\mathrm{e}})$

            $\mathbf{X}^\ddagger \leftarrow \mathrm{1layerConv1d}(\mathbf{H}')$

            Model weights updated by $\mathrm{Loss} = \mathrm{mse}(\mathbf{X}^\dagger, \mathbf{X}^\ddagger) + \left(1 - \cos(\mathbf{X}^\dagger, \mathbf{X}^\ddagger)\right)$

        **end for**

    **end for**

  **Return** Model

**Model inference and dIF calculation**

  **for** $w = 1;\ w \leq k$ **do**

    $\mathbf{X}'(w) \leftarrow \mathrm{Model}(\mathbf{X}(w))$

    Calculate $\Delta(\mathbf{X}(w), \mathbf{X}'(w))$

    **for** $i = 1;\ i \leq d$ **do**

        $\widetilde{\mathbf{X}}_i(w) \leftarrow$ permute the time course of the $i^{\mathrm{th}}$ process in random order

        $\widetilde{\mathbf{X}}'_i(w) \leftarrow \mathrm{Model}\left(\widetilde{\mathbf{X}}_i(w)\right)$

        Calculate $\Delta\left(\mathbf{X}(w), \widetilde{\mathbf{X}}'_i(w)\right)$ by Equation 2

        Calculate $\boldsymbol{c}_i(w) = \{c_{i1}(w), c_{i2}(w), \ldots, c_{id}(w)\}$ by Equation 3

        **for** $j = 1;\ j \leq d$ **do**

            $\mathrm{dIF}_{ij}(w) = \frac{1}{4}\left(\frac{1 - e^{-\lambda \times c_{ij}(w)}}{1 + e^{-\lambda \times c_{ij}(w)}}\right) \in \mathbb{R}$

        **end for**

        **Return** $\mathrm{dIF}_{i\bullet}(w) \in \mathbb{R}^d$

    **end for**

  **end for**

**Figure 3.** InfloFlowNet algorithm.

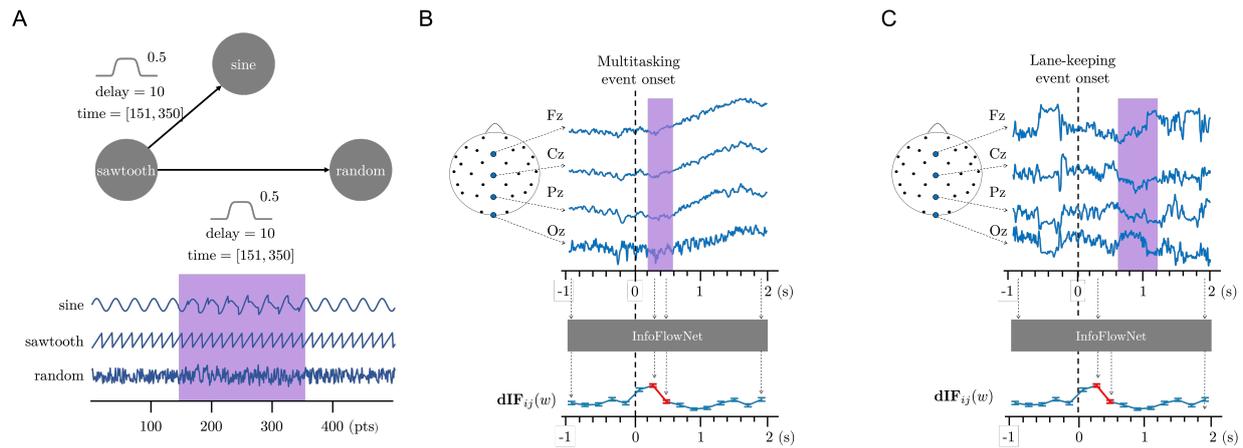

**Figure 4**. Experiment data. (A) Synthetic time-series processes, including sine, sawtooth, and random waves, and simulated causality. EEG signals collected from the (B) psychological vigilance task and (C) sustained-attention driving experiment. Color patches show the WOIs used to build the InfoFlowNet model.

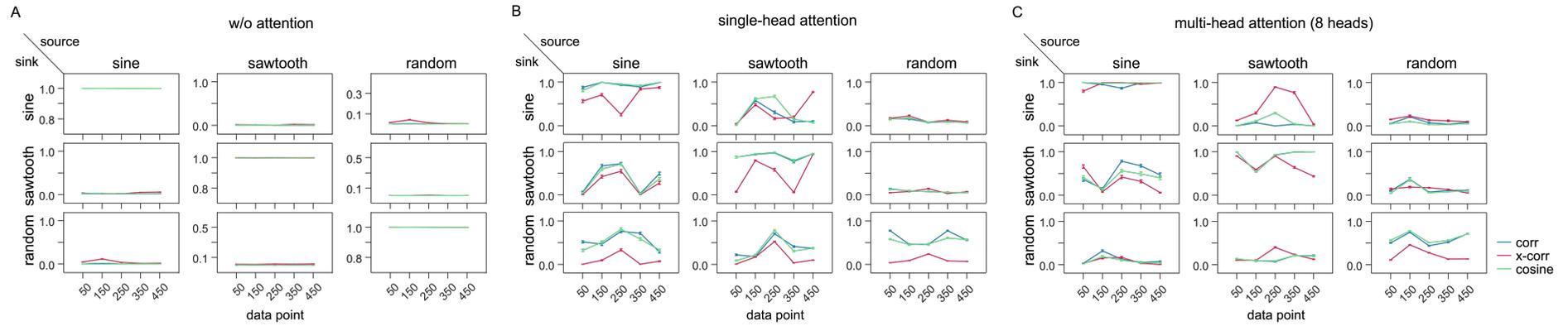

**Figure 5.** Estimation of causality magnitudes (**dIF**) between simulated source and sink processes using InfoFlowNet: (A) without attention mechanism, (B) with single-head attention, and (C) with multi-head attention mechanisms.

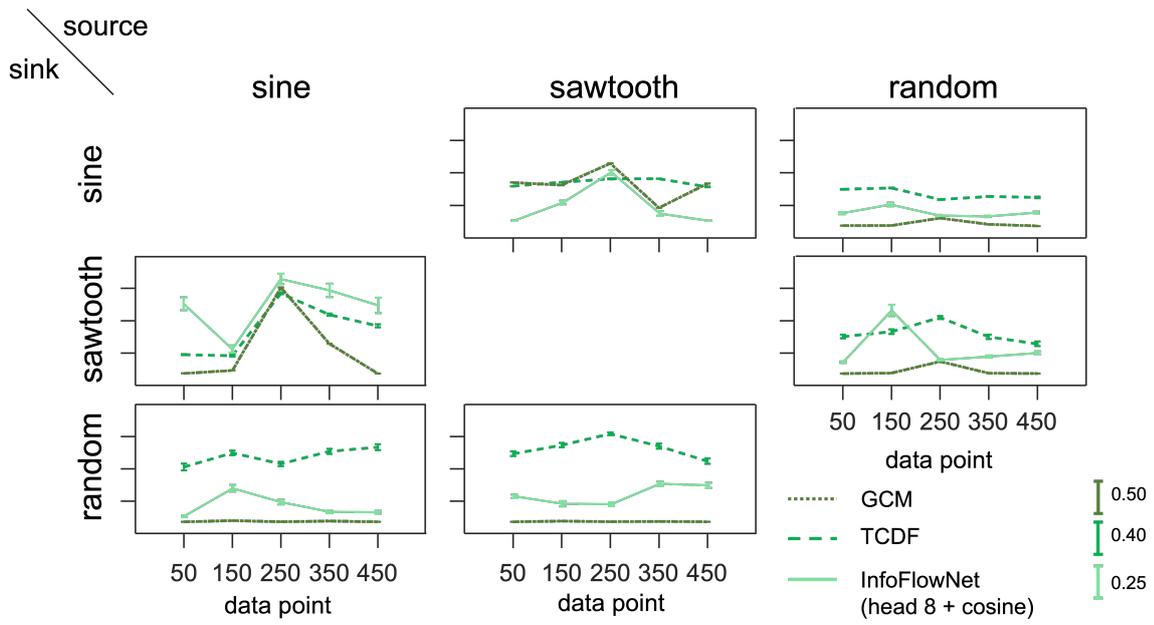

**Figure 6.** Time-varying causality changes estimated by the GCM, the TCDF, and InfoFlowNet (using multi-head attention with cosine similarity). Different scales are used for the three methods for clarity.

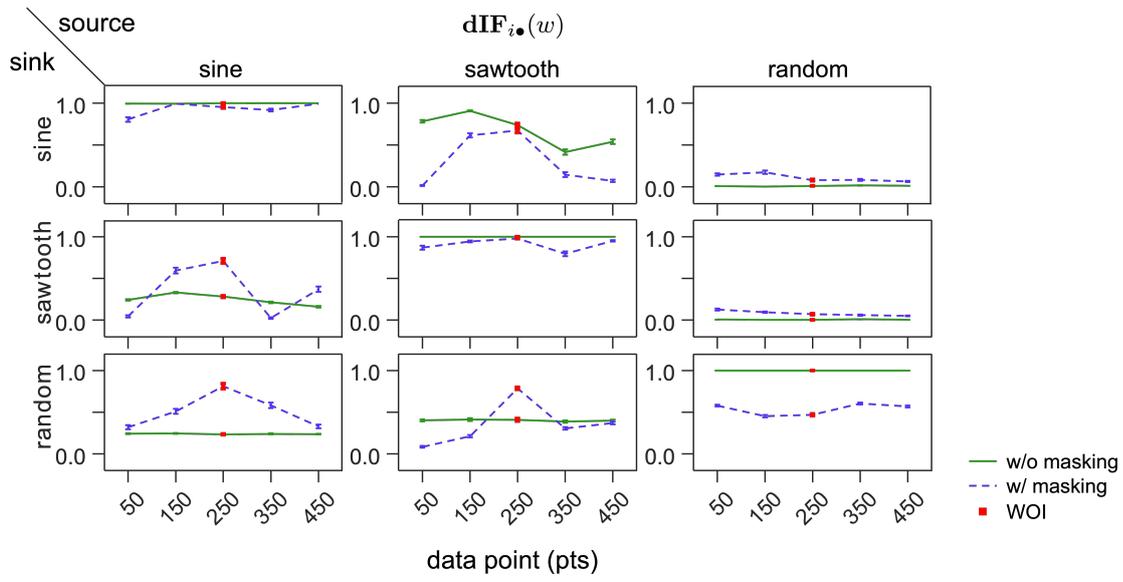

**Figure 7.** Comparison of InfoFlowNet results in simulated processes with and without masking mechanisms, using cosine similarity.

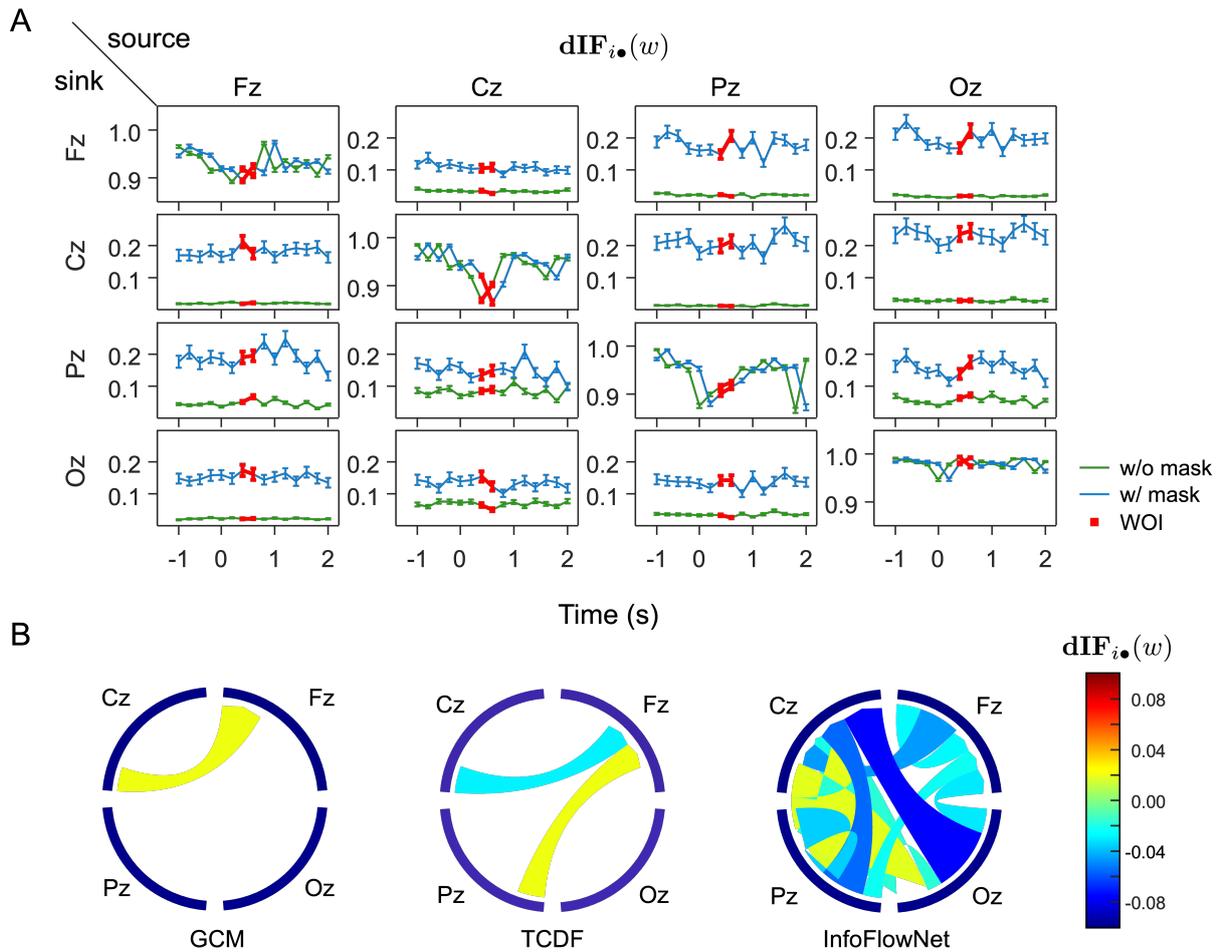

**Figure 8.** Evaluation results of the causality between four EEG signals using InfoFlowNet. (A) Results obtained by InfoFlowNet with eight heads, with and without masking and using cosine similarity. (B) Comparative connectivity graphs estimated by InfoFlowNet, the GCM, and the TCDF. The color of each edge represents the difference in causal strength before and after the event.

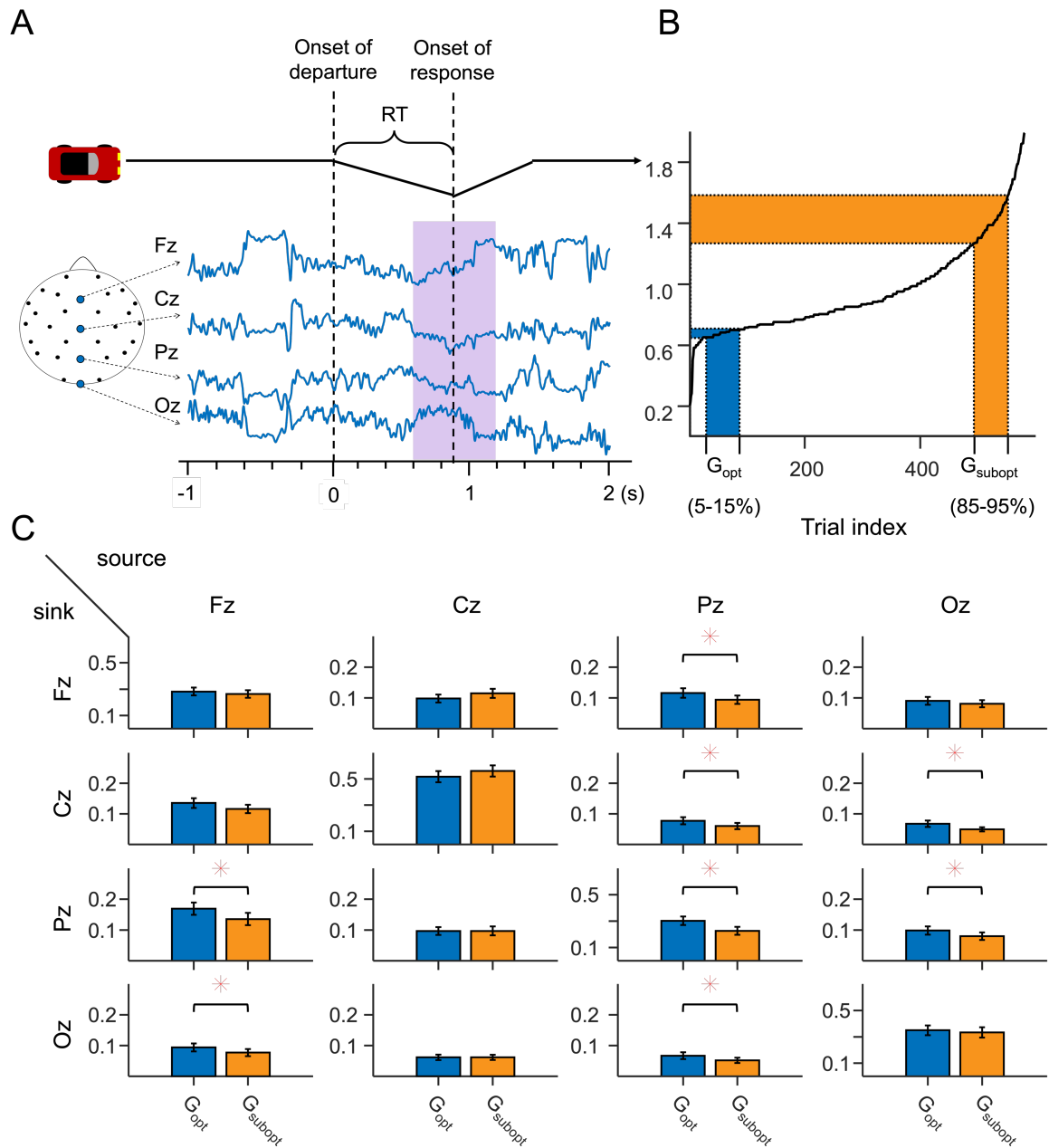

**Figure 9.** Analysis of causal relationships between four-channel processes in the lane-keeping experiment using InfoFlowNet. (A) WOI (purple), with the baseline period set from -1 to 0 s. Reaction time (RT) refers to the duration between the onset of lane departure and participant's response. (B) RT-sorted experimental trials. Trials with RTs in the 5%–15% and 85%–95% ranges are defined as $G_{opt}$ and $G_{subopt}$, respectively. (C) Comparative analysis of dIF between $G_{opt}$ and $G_{subopt}$, with asterisks indicating significant differences at a 0.05 significance level.

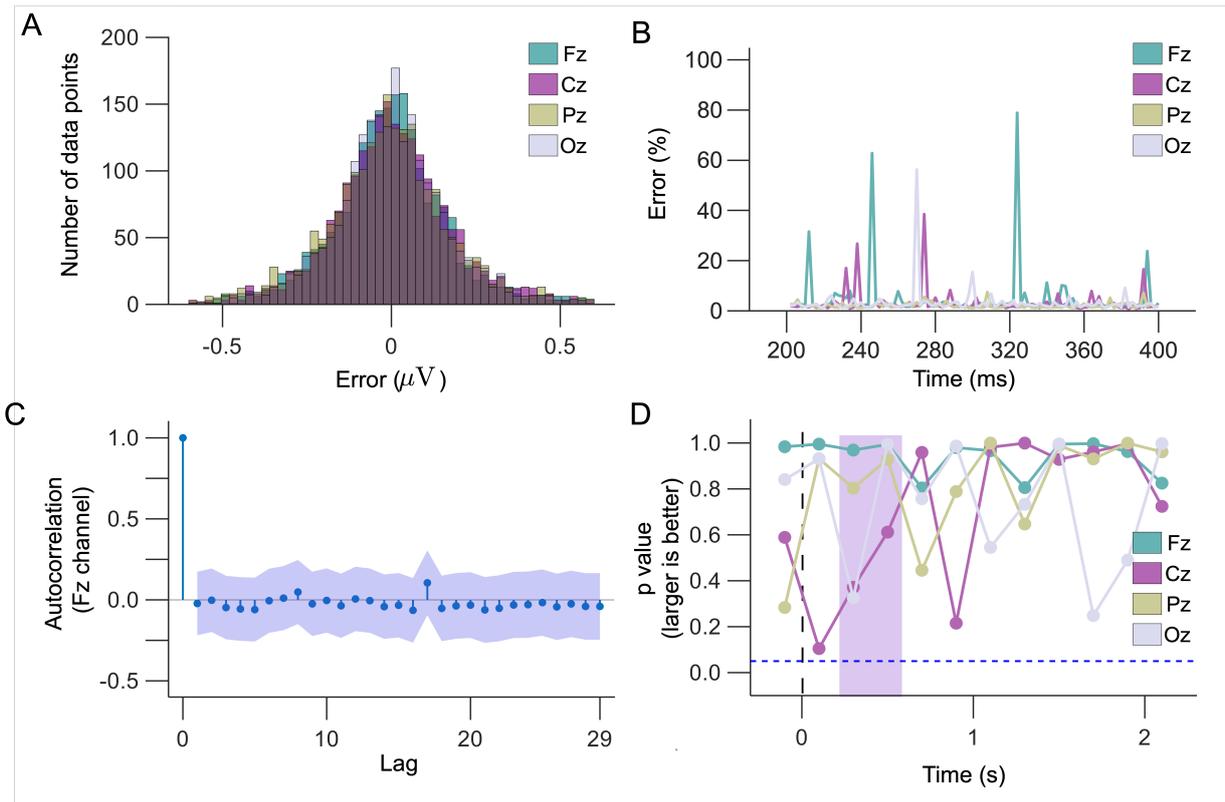

**Figure 10.** Assessment of model error whiteness. (A) Distribution of model prediction errors. (B) Percentage of prediction errors, derived from the absolute discrepancy between the actual and predicted values and then normalized by the actual value. (C) Autocorrelation function plot of prediction errors for the Fz channel within the WOI. (D) p-values obtained from Ljung–Box tests, evaluating the prediction errors across four selected channels throughout all time windows.

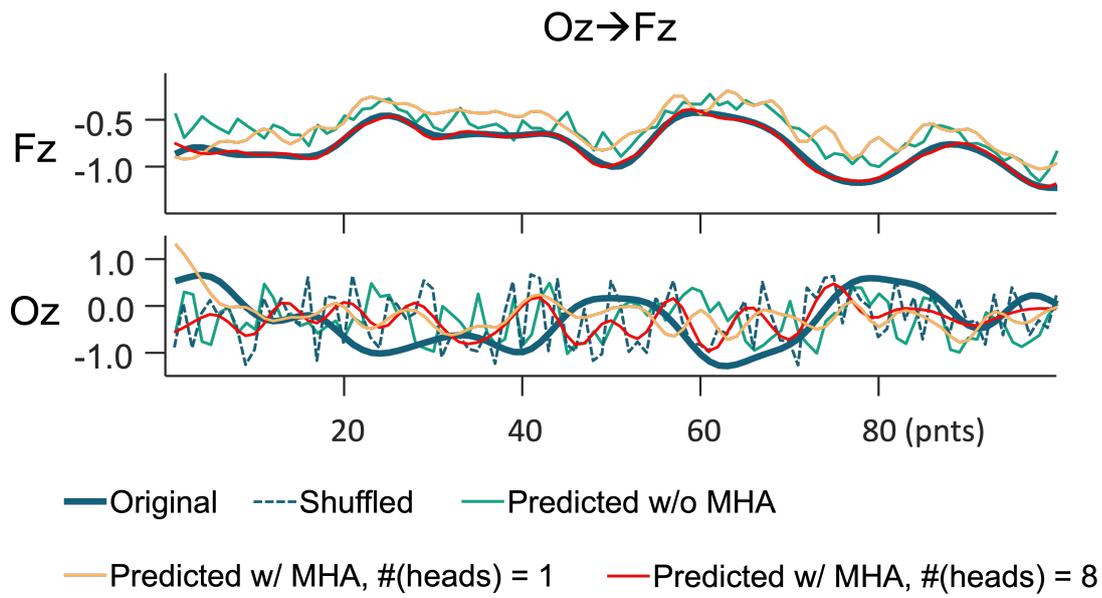

**Figure 11.** Differences between predicted and original processes with various attention configurations: without attention (w/o MHA), with single-head attention (w/ MHA, #[heads]=1), and with multi-head attention (w/ MHA, #[heads] = 8).

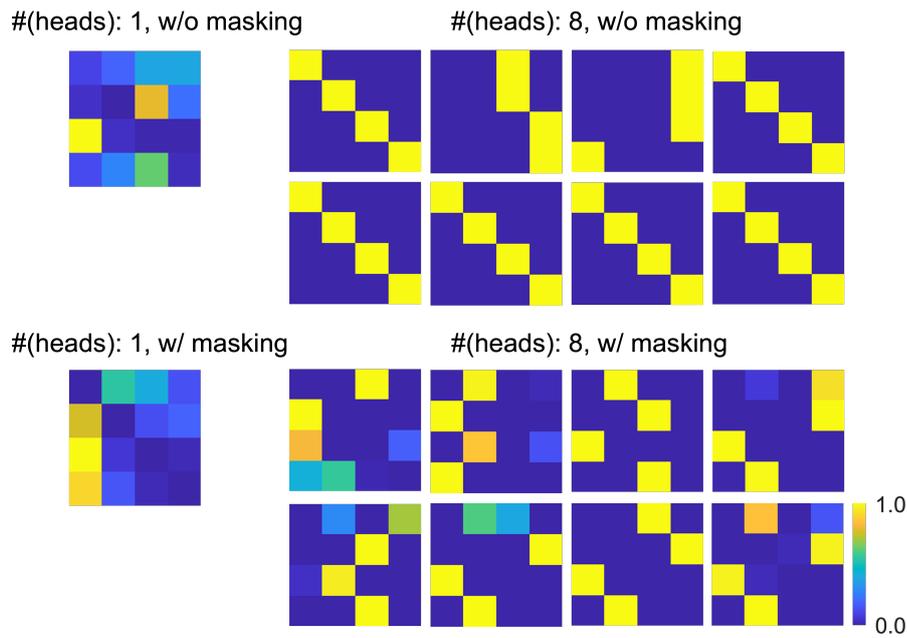

**Figure 12.** Attention score matrices from InfoFlowNet. Top-left: single head without masking. Bottom-left: single head with masking. Top-right: eight heads without masking. Bottom-right: eight heads with masking

Table 1. Computation times (in seconds) of GCM, TCDF, and InfoFlowNet

| Methods | Training | Inference | |
|---|---|---|---|
| GCM | | $5.33 \pm 0.30$ | |
| TCDF | | $355.01 \pm 1.01$ | |
| InfoFlowNet #(heads) = 1 | $118.74 \pm 10.15$ | correlation | $0.18 \pm 0.03$ |
| | | x-corr | $0.19 \pm 0.03$ |
| | | cosine | $0.27 \pm 0.04$ |
| InfoFlowNet #(heads) = 8 | $109.03 \pm 11.98$ | correlation | $0.18 \pm 0.03$ |
| | | x-corr | $0.20 \pm 0.03$ |
| | | cosine | $0.28 \pm 0.04$ |

The computation time was calculated for a dataset with a dimensionality of 4 channels $\times$ 100 points $\times$ 100 trials.